# Coherent Electron Scattering Captured by an Attosecond Quantum Stroboscope

J. Mauritsson[1], P. Johnsson[1], E. Gustafsson[1], M. Swoboda[1], T. Ruchon[1], A. L'Huillier[1] & K. J. Schafer[2]

[1]Department of Physics, Lund University, P. O. Box 118, SE-221 00 Lund, Sweden

[2]Department of Physics and Astronomy, Louisiana State University, Baton Rouge, Louisiana 70803-4001, USA

**The basic properties of atoms, molecules and solids are governed by electron dynamics which take place on extremely short time scales. To measure and control these dynamics therefore requires ultrafast sources of radiation combined with efficient detection techniques. The generation of extreme ultraviolet (XUV) attosecond (1 as = $10^{-18}$ s) pulses[1,2] has, for the first time, made direct measurements of electron dynamics possible. Nevertheless, while various applications of attosecond pulses have been demonstrated experimentally[3-5], no one has yet captured or controlled the full three dimensional motion of an electron on an attosecond time scale. Here we demonstrate an attosecond quantum stroboscope capable of guiding and imaging electron motion on a sub-femtosecond (1 fs = $10^{-15}$ s) time scale. It is based on a sequence of identical attosecond pulses[6] which are synchronized with a guiding laser field. The pulse to pulse separation in the train is tailored to exactly match an optical cycle of the laser field and the electron momentum distributions are detected with a velocity map imaging spectrometer (VMIS)[7,8]. This technique has enabled us to guide ionized electrons back to their parent ion and image the scattering event. We envision that coherent electron scattering from atoms, molecules and surfaces captured by the attosecond quantum stroboscope will complement more traditional scattering techniques[9-11] since it provides high temporal as well as spatial resolution.**

Pioneering experiments with femtosecond infrared (IR) laser fields have demonstrated that temporally localized electron wave packets (EWPs) can be used to study molecular structures and dynamics[12-14]. In these experiments the EWPs are generated through tunnel ionization twice per optical cycle near the peaks in the laser's oscillating electric field. They are subsequently accelerated by the same laser field and may be driven back to their parent ion for further interaction. If, for example, aligned molecules are used as targets their orbitals can be characterized from the resulting harmonic emission. The basic sequence of events, which is the essence of strong field physics, is very versatile and leads to many different phenomena[15,16]. The only control knob in these experiments, however, is typically the laser intensity, which must be quite high in order that there be a reasonable probability of tunneling through the Coulomb barrier. Further control of the electron dynamics requires that the creation and acceleration of the EWPs are decoupled; and this is not possible using tunneling ionization since the same laser field governs both events. Decoupling can be achieved using XUV attosecond pulses to create temporally localized EWPs through single photon ionization at a well defined phase of a synchronized IR field. These attosecond EWPs are distinctly different from tunnel ionization EWPs: they are born at the centre of the potential well with a non-zero velocity and their subsequent dynamics can be controlled by choosing the phase and amplitude of a synchronized IR field appropriately. In particular, the laser field needed to drive these EWPs back to the potential is usually far weaker than the laser field needed to form tunnel EWPs, leading to much less distortion of the electronic or nuclear properties to be studied.

Decoupling laser-driven dynamics from ionization is at the heart of the attosecond quantum stroboscope, a device to guide and image electron motion on a sub-femtosecond time scale, which is illustrated in Fig. 1a, b. The stroboscope technique is based on a sequence of identical attosecond pulses that are used to release electrons into a moderately strong laser field exactly once per laser cycle. Just as a conventional stroboscope can be used to freeze the beating of a hummingbird's wings, revealing details that would normally be blurred, we use the quantum stroboscope to record the electron momentum distribution from a single ionization event, free from interference effects due to multiple ionization events. Operationally, the quantum stroboscope works because ionized electrons receive a momentum impulse from the IR field along its polarization direction (the vertical in Fig. 1). Since this impulse depends on the magnitude and direction of the laser field at the moment of ionization[17-19], each phase of the oscillating laser field yields a unique final momentum distribution. If ionization occurs over the whole IR cycle, or even at as few as two times during the cycle, the distribution will be smeared out and show interference fringes that depend on the different ionization times[20]. When the attosecond pulse periodicity matches the optical cycle, the XUV pulses create identical EWPs that add up coherently, with the result that the properties of an individual EWP can be studied stroboscopically. In addition, the temporal localization of the EWPs within the optical period of the IR field allows for precise guiding of the wave packets after their birth.

In a first experimental demonstration of the quantum stroboscope we use pulses with a 300 as duration and a central energy of 24 eV to ionize argon in the presence of a guiding IR laser field with an intensity of $5 \cdot 10^{12}$ W/cm$^2$. The attosecond pulse train (APT) was generated from a two colour laser field consisting of the IR field and its second harmonic to ensure that the XUV pulses were separated by one full optical cycle[6]. Four stroboscopic images taken at different XUV-IR delays are presented in Fig. 1 (c). The clear up/down-asymmetry in the momentum distributions confirms that each image corresponds to ionization at one particular phase of the IR field, so that the total momentum is shifted up or down in the direction of polarization of the IR field. These results illustrate two essential aspects of stroboscopic imaging. First, we can freeze the periodically varying momentum distribution at a single phase of the IR field, and then capture the entire time-dependent distribution by varying the XUV-IR delay. Second, the measured signal is larger than what we would measure with a single pulse. In a quantum stroboscope this effect is enhanced because a train of $N$ pulses yields fringes that are $N^2$ times brighter than the signal from an isolated pulse, owing to the coherence of the process ( $N \approx 10$ in our experiment). This has allowed us to obtain full three dimensional images of an attosecond EWP for the first time. An additional feature is that the position of the stationary interference fringes depends only on the IR intensity. The quantum stroboscope is therefore self-calibrating since the only unknown parameter, the IR intensity, can be read directly from the interference pattern.

In the experimental results presented in Fig. 1 coherent electron scattering was not observed since the intensity of the guiding laser field was not high enough to direct the EWPs back to the ion core. To estimate the field intensity needed for coherent electron scattering, simple classical arguments can be used. The momentum at time t of an electron born in the field $E(t) = E_0 \sin(\omega t)$ at time $t_0$ with initial momentum $p_0$ is equal to

$$p(t,t_0) = p_0 + \frac{eE_0}{\omega}\left[\cos(\omega t_0) - \cos(\omega t)\right]. \qquad (1)$$

It is the sum of a drift momentum (which is a combination of the initial momentum obtained from the XUV ionization and of that gained in the laser field) and an oscillating term describing wiggle motion in the field. Introducing the wiggle energy $U_p = e^2 E_0^2 / 4m\omega^2$ and the initial kinetic energy $W = p_0^2 / 2m = E_{XUV} - I_P$, where $I_p$ is the ionization energy, the condition for the final (drift) momentum to be zero or opposite to the initial momentum can be formulated as $\tilde{\gamma} = \sqrt{W/2U_p} \leq 1$. When $\tilde{\gamma} = 1$, the momentum transferred by the field to the electrons is such that only electrons that are born exactly at times when $E(t)=0$ will return to the ion since the net transfer of momentum from the laser field is maximized for these times. For smaller $\tilde{\gamma}$-values the momentum transfer is larger, and returns are possible also for other initial times. We can tune the interaction (the energy of the returning electrons, the number of times they pass the core and the field strength at the moment of re-collision) by varying the delay between the attosecond pulse and the laser field. For an 800 nm laser wavelength, $\tilde{\gamma} = 1$ can be obtained by using, for example, $I = 1 \cdot 10^{13}$ W/cm$^2$ and W=1.2 eV. This intensity is an order of magnitude smaller than that needed for tunnel ionization.

To understand what we can expect from experiments that guide ionized electrons to rescatter off the ion core, we have performed a series of calculations in helium by numerically integrating the time-dependent Schrödinger equation (TDSE). The momentum distributions obtained at the XUV-IR delay leading to maximum momentum transfer from the guiding field to the electrons are shown in Fig. 2 for four different IR intensities. The white circle in each panel denotes the range of momenta expected if there is no rescattering from the ion. We clearly see the onset of scattering for $\tilde{\gamma}$ values near one ($\tilde{\gamma} \approx 1.2$). At this intensity the low energy fraction of the EWPs has scattered off the atomic potential while the high energy portion remains unaffected. With decreasing $\tilde{\gamma}$ values a larger portion of the EWP scatters off the potential, which increases the scattering signal.

To experimentally access the regime of coherent electron scattering we change the target gas to helium, so that W=1.2 eV, and we increase the IR intensity to $I = 1.2 \cdot 10^{13}$ W/cm$^2$. Four experimental momentum distributions recorded at different XUV-IR delays are presented in Fig. 3. When the XUV-IR timing is set to maximize the momentum transfer from the IR field in the upwards (panel 1) or downwards (panel 3) directions we see a clear signature of re-scattering, manifested by a significant increase of low-energy electrons in the direction opposite to the momentum transfer from the IR field[21]. The experimental results are compared with theoretical calculations in Figure 4 and the agreement is excellent with all the substructures well reproduced. We believe that this is the first evidence for coherent electron scattering of attosecond EWPs created by single photon ionization.

In this letter we have demonstrated an attosecond quantum stroboscope capable of imaging the electron momentum distribution resulting from a single ionization event. We have also used it to guide ionized electrons back to their parent ions and to image the coherent electron scattering. The basic technique we have demonstrated is very versatile and may be altered in a number of potentially useful ways. For example, the guiding field could be a replica of the two-colour driving field we used to make the attosecond pulses. This would provide additional control over the return time and energy of the electrons. Using a longer wavelength driving laser would lengthen the time between attosecond pulses, allowing more time for internal dynamics initiated by the launch of the EWP to develop before being probed by the returning electron. We envision that controlled, coherent scattering such as we have demonstrated will enable time resolved measurements with very high spatial resolution in atoms and molecules or at surfaces.

**Acknowledgments:** This research was supported by a Marie Curie Intra-European Fellowship, the Marie Curie Research Training Networks XTRA, the Marie Curie Early Stage Training Site MAXLAS, the Integrated Initiative of Infrastructure LASERLAB-EUROPE within the 6th European Community Framework Programme, the Knut and Alice Wallenberg Foundation, the Crafoord foundation, the Swedish Research Council and the National Science Foundation.

**Author Information** The authors declare no competing financial interests. Correspondence and requests for materials should be addressed to J.M. (johan.mauritsson@fysik.lth.se).

**Figure 1: Principle of the attosecond quantum stroboscope.** An attosecond pulse train containing *N* pulses is used to ionize the target atoms once per cycle of an IR laser field. This periodicity ensures that the created EWPs are identical and that they add up coherently on the detector leading to an enhanced signal that is modulated by interferences fringes (a,b). The bright fringes are increased by a factor $N^2$ compared to the signal that would be obtained with only one pulse. The IR field is used to guide the electrons and the momentum transfer can be controlled by simply varying the XUV-IR delay. When the EWPs are created at the maxima of the IR electric field (a) the net transfer of momentum is zero and the resulting momentum distribution is symmetric relative to the plane perpendicular to the laser polarization. When the EWPs instead are created at the zero-crossings of the IR electric field (b), the momentum distribution is shifted by the field up- or down-wards along the direction of the laser polarization. In our experiment, the final momentum distributions of the EWPs are detected using a VMIS. Experimental results obtained in Ar at four different XUV-IR delays are shown in (c), panels 1-4 correspond to the delays $t_0$=-π/2ω, 0, π/2ω, π/ω respectively for an IR intensity of $5 \cdot 10^{12}$ W/cm$^2$. The delay-dependent momentum distribution is imprinted on a set of concentric rings which are due to the coherent addition of many EWPs. These rings, which are evenly spaced one IR photon apart in energy, show a decreasing spacing when plotted as a function of momentum. The positions of the rings do not shift as a function of XUV-IR delay, which shows that we are imaging a train of identical EWPs spaced one IR cycle apart in time.

**Figure 2: Theoretical illustration of coherent scattering.** Theoretical results obtained by integrating the TDSE are shown for IR intensities ranging from zero to $2 \cdot 10^{13}$ W/cm$^2$ for the XUV-IR delay which corresponds to the maximum momentum transfer. The momentum transfer from the field to the electrons is upward in the figure. The white circles are positioned at the highest energy electrons in the field free case (panel 1) and shifted by the amount of momentum added by the IR field in the other panels. If no post-ionization interaction between the electron and the atom occurs the momentum distributions would remain within the circles. This is the case in panel 2, in which the IR field is too weak to drive the electron back to the atomic potential, but not in panel 3 and 4 where electrons appear outside the circles in the downward direction. Two features are highlighted in panel three, which is calculated at an intermediate intensity where $\tilde{\gamma}$ = 1.2, by white arrows: I) electrons that have scattered off the core appearing outside the white circle in the downward direction and II) interference minima in the momentum

distribution. The interference minima occur in the region where the rescattered and direct electrons overlap.

**Figure 3: Experimental demonstration of coherent electron scattering.** Experimental results obtained in helium at an intensity of $1.2 \cdot 10^{13}$ W/cm$^2$ at the same delays as in Fig. 1 (c) are shown. The results are distinctively different from those taken in argon (Fig. 1). The momentum distributions are more peaked along the laser polarization direction, since in helium the excited EWP is entirely in an *m*=0 state, whereas in argon there was a mixture of *m*=0 and *m*=1 and the latter has no amplitude along the polarization axis. This is a favourable condition to observe electron scattering since the electrons along the polarization direction have the highest probability to scatter off the potential. With this higher intensity more momentum is transferred to the electrons and in combination with the lower initial energy some electrons return to the atomic potential for further interaction.

**Figure 4: Comparison between theory and experiment.** We compare the experimental results (right) with theoretical calculations (left) obtained for the same conditions. The excellent agreement is the strongest evidence for coherent scattering effects in the experiment. All the substructures well reproduced except for the innermost peak in the experiment, which most likely is due to above threshold ionization of residual water in the experimental chamber and therefore not included in the theoretical results.

Figure 1

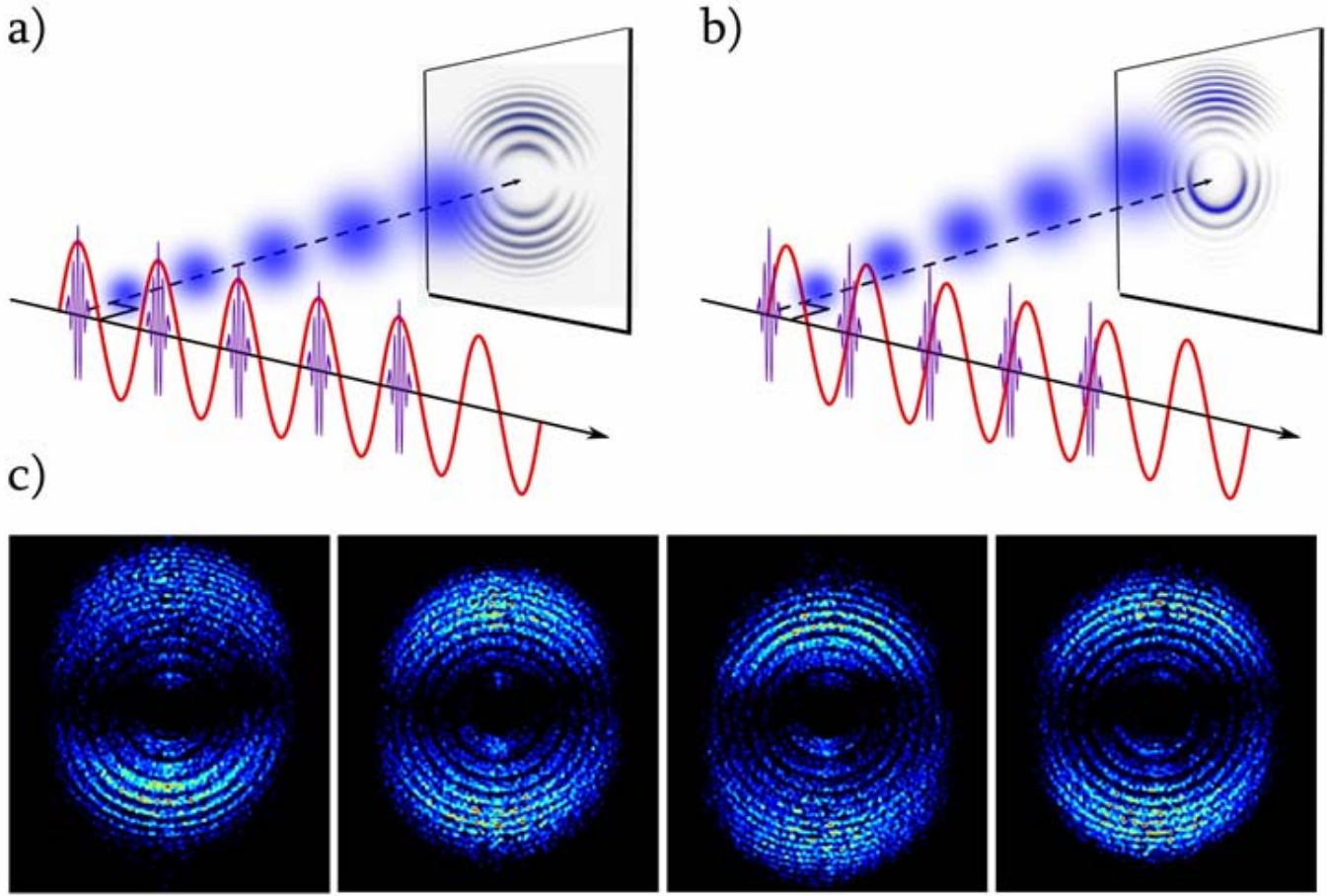

Figure 2

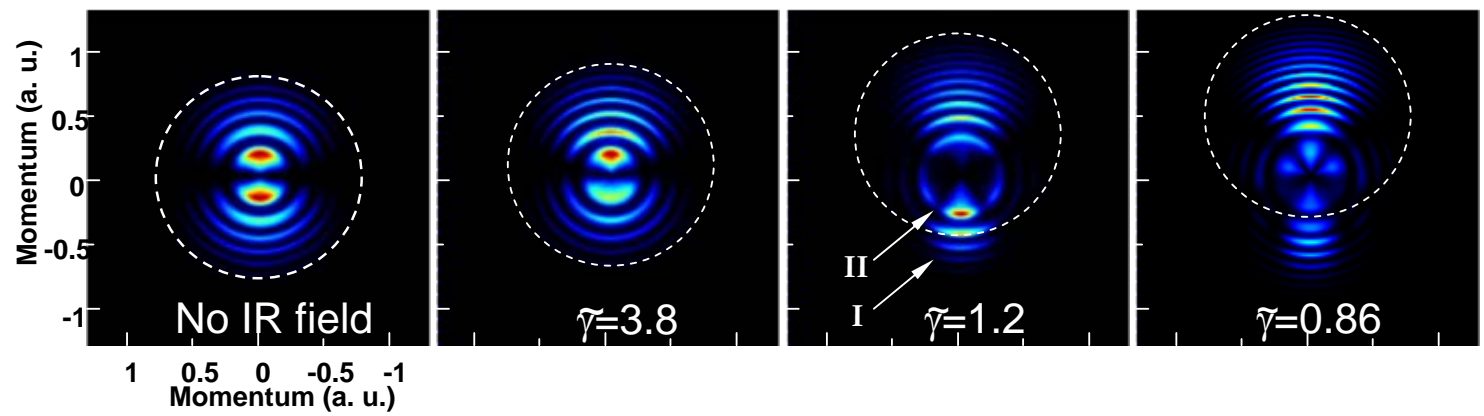

Figure 3

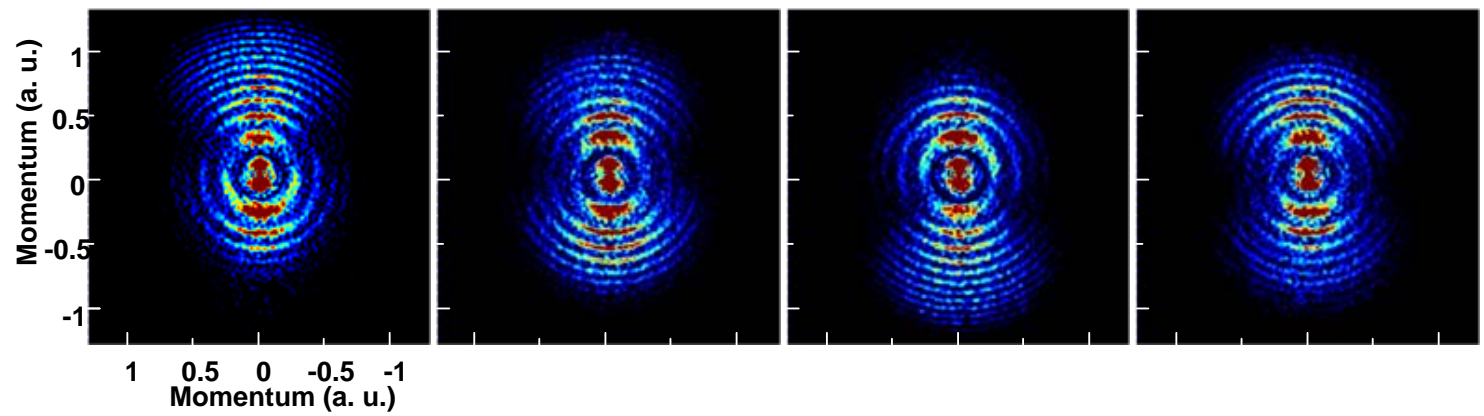

Figure 4

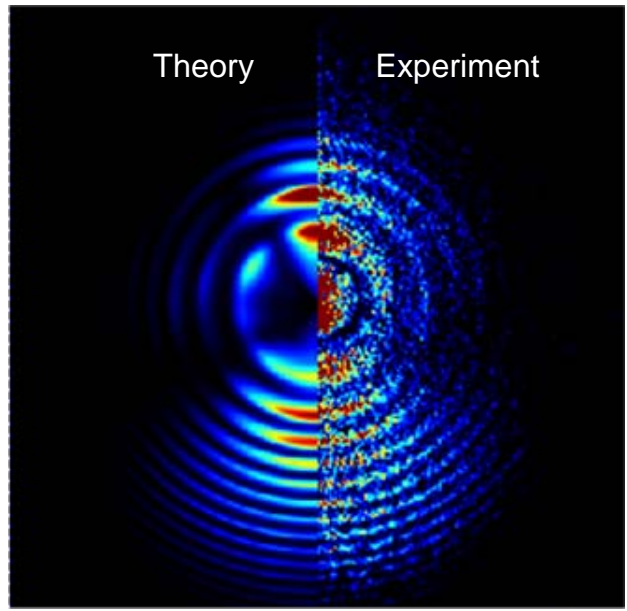